\documentclass{pasa}%

\title[Is the stellar system WR\,11 a gamma-ray source?]{Is the stellar system WR\,11 a gamma-ray source?}
\author[Paula Benaglia.]{Paula Benaglia$^{1,2}$\thanks{pbenaglia@fcaglp.unlp.edu.ar}\\
\affil{$^1$Instituto Argentino de Radioastronom\'{\i}a, CCT-La Plata, CONICET,
              Argentina}%
\affil{$^2$Facultad de Ciencias Astronómicas y Geofísicas, UNLP, Paseo del Bosque s/n, 1900, La Plata, Argentina}}%
\jid{PASA}
\doi{10.1017/pas.\the\year.xxx}
\jyear{\the\year}

\usepackage[authoryear]{natbib}
\bibpunct{(}{)}{;}{a}{}{,}
\usepackage{aas_macros}
\usepackage{hyperref} 
\hypersetup{colorlinks,citecolor=blue,linkcolor=blue,urlcolor=blue}

\begin{document}%
\begin{abstract}
Many early-type stars are in systems; some of them have been indicated as putative high-energy emitters. The radiation is expected to be produced at the region where two stellar winds collide. Compelling evidence of such emission was found only for the colliding-wind binary (CWB) Eta Car, which was associated to a GeV source. Very recently, the closest CWB, WR\,11, was proposed as a counterpart of a 6$\sigma$ emission excess, measured with the Fermi LAT satellite. We looked for evidence to support or reject the hypothesis that WR\,11 is responsible of the gamma-ray excess. 
Archive radio interferometric data at 1.4 and 2.5 GHz taken with the Australia Telescope Compact Array along sixteen different dates were reduced. The sizes of the field-of-view at 2.5 GHz and of the central region of the Fermi LAT excess are alike. We analyzed the emission of the field of WR\,11, characterized the radio sources detected and derived their spectral indices, to investigate their nature. Eight sources with fluxes above 10 mJy were detected at both frequencies. All but one (WR\,11) showed negative spectral indices. Four of them were identified with known objects, including WR\,11. A fifth source, labeled here S6, could be a promising candidate to produce gamma-ray emission, besides the CWB WR\,11. 
\end{abstract}
\maketitle%
\section{INTRODUCTION}
\label{sec:intro}

The first gamma-ray all-sky observations, obtained decades ago with the satellites COS-B \citep{1983SSRv...36...61H} and Compton \citep{1999ApJS..123...79H}, disclosed numerous sources with not known counterpart at other wavelengths, hereafter called unidentified gamma-ray sources or UNIDS. Since then, a large number of multifrequency observations have been implemented to clarify the nature of those sources 
\citep{2013ApJS..209...10M,2008A&A...482..247P}. Nowadays, though telescope capabilities have been largely expanded, like the source localization, there still remain hundreds of gamma-ray sources to be identified. For instance, the third Fermi LAT catalog \citep[][around 3000 sources, {hereafter 3FGL}]{2015ApJS..218...23A}
listed one-third of the sources with unknown counterpart at other energy range.

The identified gamma-ray sources are mostly Active Galactic Nuclei, pulsars, SNRs and High-Mass X-ray Binaries. Many of these objects emit at radio waves, especially at low frequencies, where synchrotron radiation is stronger. This turns the radio band the most preferred to search for UNIDS counterparts. 

The information collected at radio frequencies often serves to predict the behavior of the systems in the high-energy (HE) domain, and the predictions can be compared with the observations. The same population of particles (i.e. electrons) is involved  at the same time 
with magnetic fields, producing low energy synchrotron photons, as well as with photon fields, producing HE photons through Inverse-Compton (IC) scattering. 
In this way,  the radio data can be used to impose severe constraints to the HE spectrum. The results will thus help to develop more complex and accurate models, leading to a fruitful iteration between theory and observations. 

Some early-type (OB) stars and their descendants Wolf-Rayet stars have been indicated as putative high-energy emitters \citep{1999A&A...348..868R}. If two (or more) of those stars form a stellar system, their strong winds can interact with magnetic fields and relativistic particles,  in a wind-collision region, as was described by \citet{1993ApJ...402..271E}. 
Candidates, scenarios and conditions under which high-energy can be produced are given in \citet{2000MNRAS.319.1005D}, \citet{2001A&A...366..605B}, \citet{2003A&A...399.1121B}, \citet{2006ApJ...644.1118R}, 
\citet{2006A&A...446.1001P}, \citet{2009ApJ...694.1139R}.
Lately, gamma-ray AGILE satellite observations allowed \citet{2009ApJ...698L.142T},
 for the first time, 
  to present evidence of high-energy emission from the CWB Eta Car, an LBV+O? system. 
  Additional confirming evidence using Fermi LAT data \citep{2010ApJ...723..649A} was soon 
 published.   

After compiling radio, X-ray and gamma-ray data for OB and WR stars, \citet{2013A&A...558A..28D}
presented a very comprehensive catalog of particle acceleration colliding wind binaries, or PACWBs, with more than 40 candidates, and discussed the main problems related to this subject. \citet{2013A&A...555A.102W}
modeled possible HE emission produced by WR\,11 and tagged it as one of the best CWB candidates to be detected with the Fermi LAT. 
The latest Fermi LAT release reunited data from 2008 to 2015. \citet{2015A&A...577A.100R} derived from it an increase of the flux coming from Eta Car as the system approached periastron. 
\citet{2015arXiv151003885P} also analyzed the 7-yr Fermi LAT data, and reported emission excess coincident with
the field of the nearest WR+OB system, WR 11. He ascribed the emission to the stellar system. However, the Fermi LAT excess extends over a square degree -the maximum itself covers about 10 square arcmin of the sky-, encompassing hundreds of sources. And there is a significant volume of radio observations that targeted WR\,11, that cover the central region of the Fermi source. 
I have carried out the reduction and analysis of high-resolution archive {radio} data of the WR\,11 field, in looking for additional possible counterparts to the Fermi emission. The results are presented here. 

Section 2 contains the information of the system WR\,11 that is relevant to this study. Section 3 describes the archive data reduction and the stellar  field in the light of the radio continuum observations. Correlation results are presented in Sect. 4 and discussed in Sect. 5. In Section 6 the main conclusions are highlighted.

\section{THE STELLAR SYSTEM WR\,11}

The source WR\,11, also known as Gamma$^2$ Velorum or $\gamma$ Vel, is a double-lined spectroscopic binary system, and the nearest CWB, composed by a WC8 and a O7.5\,II-III stars. It is located at RA, Dec (J2000) = 8:09:31.95, $-$47:20:11.71,  at a distance of  $\sim$370 pc \citep{2007A&A...464..107M}, and has a period of 78.5 d 
\citep{2007MNRAS.377..415N}.
The object was observed with the Australia Telescope Compact Array\footnote{The Australia Telescope Compact Array is funded by the Commonwealth of Australia for operation as a National Facility managed by CSIRO.}
(ATCA)
\citep{1997ApJ...481..898L,1999ApJ...518..890C}
at four bands, from 1.4 to 8.64 GHz. The measured fluxes allowed to derive a spectral index {$\alpha$} decreasing from 1.2 to 0.3 with growing frequency. 
{(Throughout the text, we adopted the convention $S \propto \nu^{\alpha}$, where $S$ is the radio flux density at the frequency $\nu$.)}
Based on that, \citet{1999ApJ...518..890C}
stated that the emission from WR\,11 could have some non-thermal {(NT)} contribution. 
The system is listed in the PACWBs catalog, along with detailed information.
At X-rays, the source is detected and the emission is explained by a hot {collisional} plasma, the stellar winds; the low energy photons are absorbed \citep{2004A&A...422..177S}. 

Optical spectro-polarimetric data of WR\,11 with the ESPaDOnS Canada–France–Hawaii Telescope provided
no definite detection of the magnetic field in the winds of WR\,11 \citep{2014ApJ...781...73D}, but a field strength upper limit of $B_{\rm max} \sim 500$ G. 
Data collected after the first 2 yr by Fermi LAT were analyzed 
\citep{2013A&A...555A.102W}, with negative results on a WR\,11 detection. 
{The distance from WR\,11 to the closest known 3FGL source (J$0800.6-4806$, of $\sim 0.3^\circ$ semimajor axis) is 
1.7$^\circ$ in the plane of the sky.}
\citet{2015arXiv151003885P} compiled Fermi LAT data taken from 2008 Aug 4 to 2015 Jul 1. 
The author reduced them with the same latest routines as the Third Fermi catalog \citep{2015ApJS..218...23A} and galaxy emission models \citep[see also][for details on Fermi LAT data reduction]{2013A&A...555A.102W}. No point sources were discovered, but a 6.1$\sigma$ excess of gamma-ray emission was detected. The excess covered an area of 1 deg$^2$, where the binary system WR\,11 is close to the centre. The location of such excess is given in TS (test statistics) values, similar to probability contours. The maximum contour of 37 TS (=6.1$\sigma$) extends in a circle of 3 arcmin diameter, and encompasses WR\,11 \citep[see Fig. 3 of][]{2015arXiv151003885P}.

A search of the Simbad database in a 15 arcmin radius circle centred at WR\,11 resulted  in $\sim$ 470 objects, including Young Stellar Objects and protostellar candidates ($\sim$200), stars ($\sim$160) and X-ray sources ($\sim$60).
WR\,11 belongs to the Vela OB2 association that gathers a hundred early-type candidate members (de Zeeuw et al. 1999).
\citet{2000MNRAS.313L..23P}
identified an association of low-mass pre-main sequence stars  surrounding WR\,11,  showing strong X-ray emission. 
 Many of the stars are assembled in the Gamma Vel cluster \citep{2014A&A...563A..94J}.
 A Spitzer survey for circumstellar
material around the low-mass association members reported by \citet{2008ApJ...686.1195H} revealed a low disc frequency.

\section{THE FIELD OF WR\,11 AT RADIO CONTINUUM}

I searched the Australia Telescope Online Archive (ATOA, www.atnf.csiro.au/atoa) for data at the lowest frequency bands, L (1.4 GHz) and S (2.5 GHz), where non-thermal (synchrotron) emission should be stronger\footnote{Analysis of archive data at higher frequencies will be given elsewhere.}. The field of WR\,11 was observed from 1999 to 2001, at various array configurations including 6*, 375, 1.5*, 750* and EW352 (Project C787), often during LST ranges of a few hours. 
Raw data taken along fifteen different days from Jun-12 to Dec-12 2001 provided enough sensitivity and resolution to detect WR\,11 and were reduced for the present study. {The configuration set comprised baselines between 30 and 6000 m; the sizes of the largest well-imaged structures were 8' and 13' at 2.5 and 1.4 GHz\footnote{
 http://www.narrabri.atnf.csiro.au/observing/users\_guide/users \_guide.html}.} 
The observing bands, of 128 MHz bandwidths, were centred at 1.384 MHz and 2.496 MHz. The FoVs were $33'$ and $20'$ respectively. The total time on source was $\sim$ 12 h at each band.
The source 1934$-$638 served as flux calibrator and the source 0823$-$500 was monitored as phase calibrator\footnote{The archive data used correspond to the ATCA configurations 6A, 6B, 6D, 1.5A, 1.5D, 750D, 375 and EW352. The observing dates were 6-Jun, 14-Jun, 1-Jul, 4-Jul,  31-Jul, 2-Aug, 24-Aug, 28-Aug, 5-Sep, 22-Sep, 13-Oct, 3-Nov, 26-Nov and 12-Dec 2001.}.

The {\sc miriad} package was used for data reduction, which was performed in the standard way. The best images were obtained with robust weighting, and a rms near the theoretical one at each band was attained. At the image centre, the resulting rms remained below 0.1 mJy beam$^{-1}$ at each band. The synthesized beams configured at highest resolution were $11.8'' \times 2.7''$, $PA=45.9^\circ$ (S band) and $22.6'' \times 5.8''$, $PA=43.6^\circ$ (L band). The beams are very elongated because the LST ranges at most of the configurations were short (2 to 3 h), since the primary goal of project C787 was not to give the best image of the field, but to focus on WR\,11. The highest-resolution images were convolved with circular beams of $24'' \times 24''$ (L band) and $12'' \times 12''$ (S band) (Fig. 1).

The area to analize was constrained by and thus circumscribed to the extension of the radio image at the larger frequency (given by its primary beam), but it agreed also with the central region of the Fermi LAT excess reported by \citet{2015arXiv151003885P}: see Figure 1, left.

The ATCA data revealed no extended emission. The L-band image showed many more sources than the S-band one, at the same sky area. 
The {search} for HE counterparts was focused on the study of the stronger sources at L band. We selected the objects detected with an integrated flux of 10 mJy or above. These values are of the order of those usually quoted for radio-gamma SED fits \citep[e.g.][]{2012A&A...543A..56D}.
Eight sources fulfilled the mentioned condition. They are listed in Table 1, and labeled S1 to S8, in order of increasing Right Ascension. 

S3, S5 and S7 are discrete sources at both bands\footnote{The deconvolved major and minor axis after a Gaussian fit resulted $7.1'' \times 4.7''$ for S3, $4.1'' \times 1.7''$ for S5, and $7.3'' \times 2.5''$ for S7.}. S1, S4 and S6, two-peaked, appeared as double sources. S8 is  slightly elongated to the east, and S2 presents some substructure. 

The integrated fluxes were derived either by fitting Gaussians to each discrete source or by measuring the flux above the 3$\sigma$ contour (1$\sigma$ = 1 rms), in the higher resolution images. In the latter case, the errors quoted in the flux densities correspond to the difference between the flux above 3$\sigma$ and that above 2$\sigma$. 

Spectral indices were computed from the integrated flux at 1.4 GHz and 2.5 GHz for each source, as $\alpha = \log (S_2/S_1)/\log (\nu_2/\nu_1)$, and are listed in Table 1.

The system WR\,11 had been also observed in Feb 1998 (C599) at S and L bands, at configuration 6A, 128-MHz bandwidths and same calibrators as for C787 data. I reduced the raw data corresponding to 15 h on source taken during day 27, to obtain images with the highest angular resolution possible with the ATCA and as an additional check for flux measurements and spectral index values. The synthesized beams, with robust weighting, resulted in $10'' \times 7''$ (L band) and $5'' \times 3''$ (S band), and the rms equal to 0.06 mJy beam$^{-1}$. 

The WR\,11 field characterization was completed with the survey of the sources above 3 times the larger noise at the L-band image ($\sigma$ = 0.2 mJy beam$^{-1}$). Table 2 lists all the sources -discrete and point-like- discovered in that way. Their fluxes were obtained by Gaussian fitting and the Simbad database was searched for possible counterpart identifications (see Table 2).

   \begin{figure*}
   \begin{center}
   \minipage{0.48\textwidth}%
   \includegraphics[width=\textwidth]{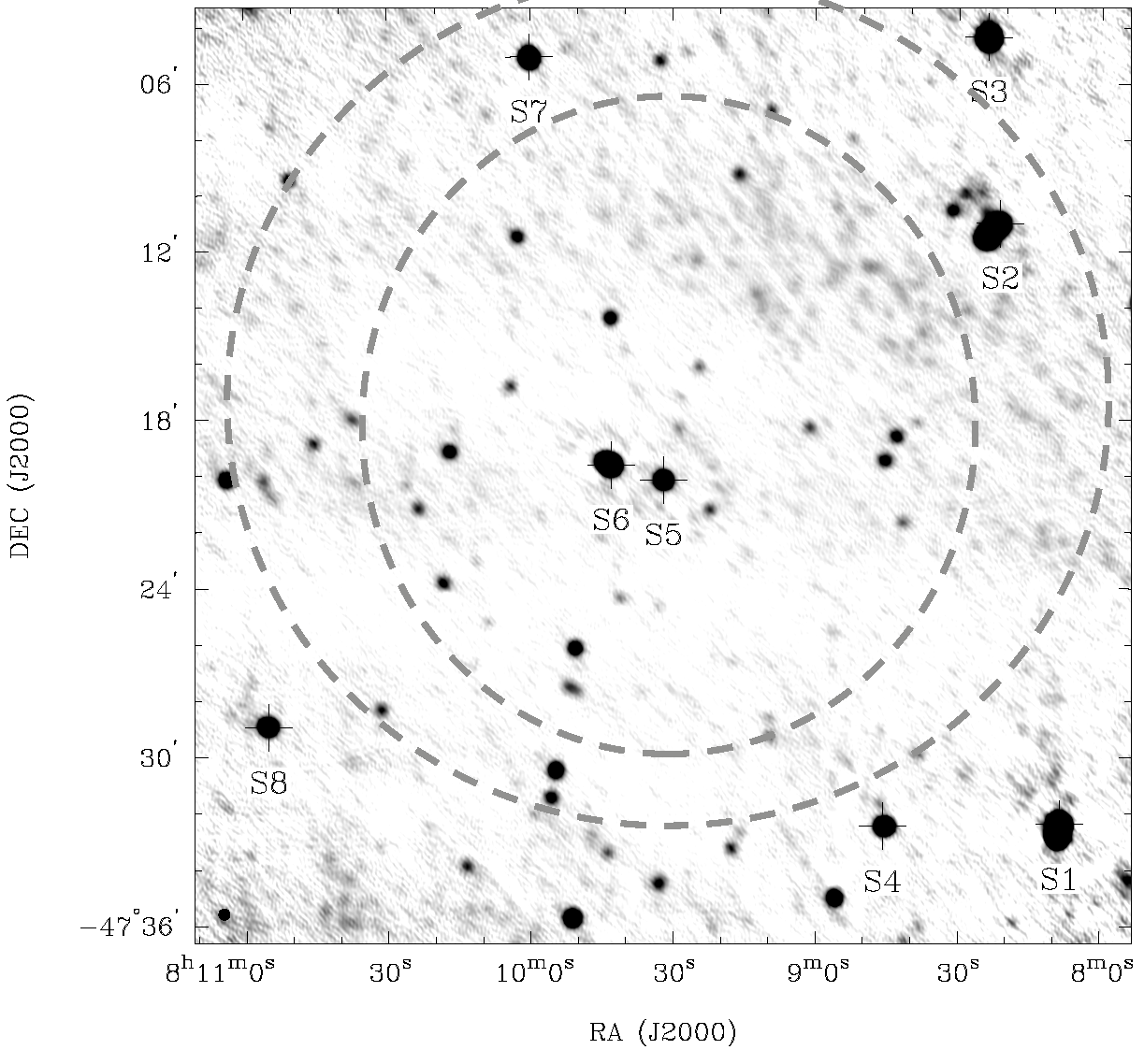}
   \endminipage\hfill
\minipage{0.48\textwidth}%
      \includegraphics[width=\textwidth]{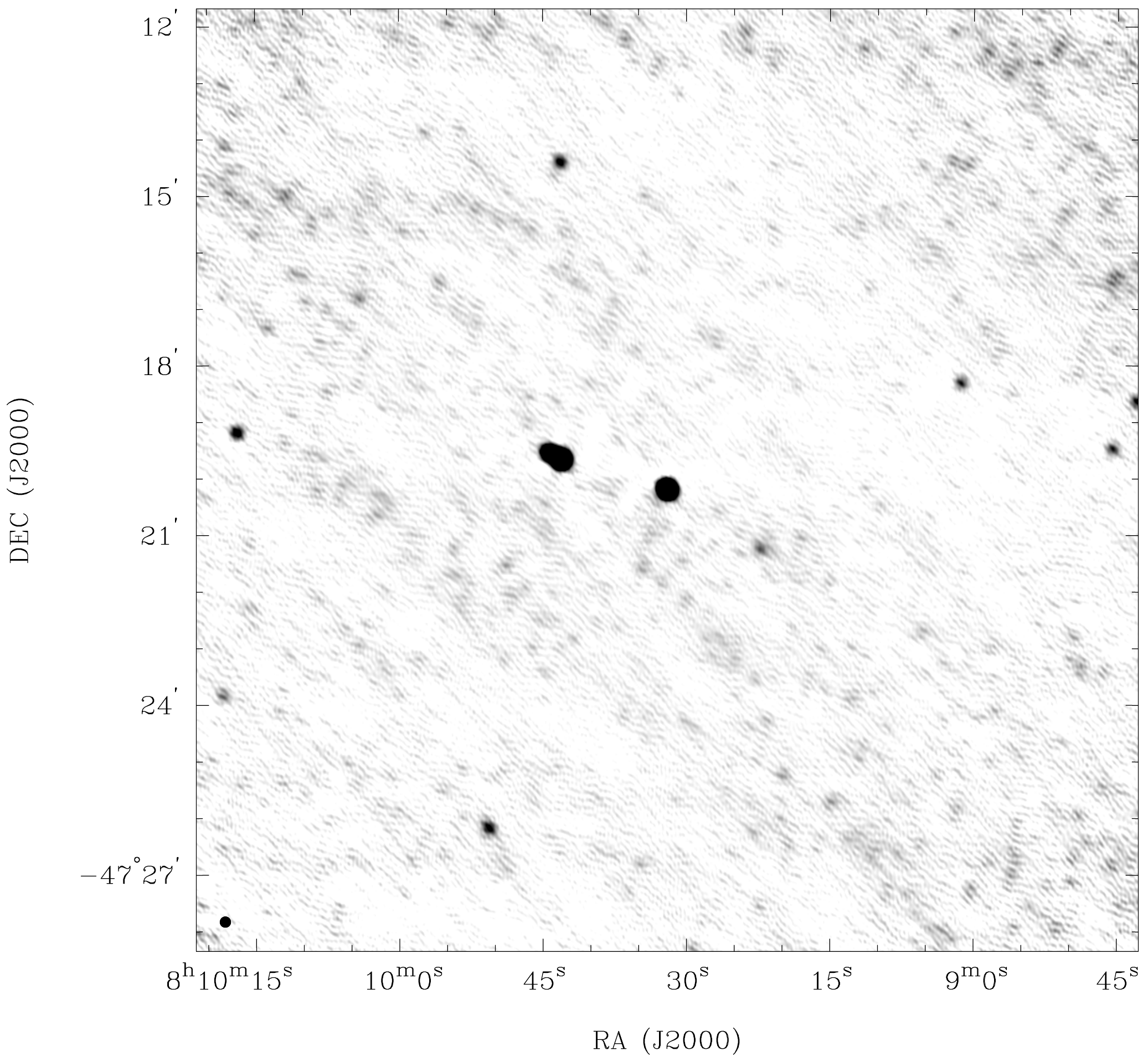}
   \endminipage
      \caption{Radio continuum emission at 1.384 GHz (left) and 2.496 GHz (right), central regions; ATOA data (project C787). The rms of each map is $\leq$0.2 mJy beam$^{-1}$. The synthesized beams are 24'' x 24'' and 12'' x 12''. The brightest sources are marked with crosses and named S1 to S8 (see also Table 1). The dashed circles in the left image represent the Test Statistics (TS) at 20 (outer) and 25 (inner) of the Fermi LAT excess 
 \citep[see {Fig. 2 of}][for details]{2015arXiv151003885P,2013A&A...555A.102W} .}    
 \label{atoadata}
 \end{center}
   \end{figure*}

\section{RESULTS}

Helped with Simbad and NED databases, we carried out a search of objects in 4-arcmin radius circles around sources S1 to S8, to seek information about  their nature. 
{Such search radius granted us to collect sources from web lists at the different spectral ranges, and not only the recent most accurate ones; in the case of an extended source-counterpart, we sought its extension in the literature.
Figures 2 and 3 display} the individual images of S1 to S8 at maximum angular resolution (data from project C599) at one frequency; the objects found at the web databases in each sub-field are also represented. 

The source S1 appears as double at both observing bands. Its spectral index is negative ($-0.6\pm0.14$). \citet{2008ApJ...686.1195H} listed three Young Stellar Objects superposed to S1, and named them as \#74 (southern one), \#78 (central one) and \#81 (northern one), see Table 1 and Fig. 2. According to IR (Spitzer) colors, the objects 74 and 81 are characterized as diskless; and the central object, \#78, corresponds to a proplyd: an evolved disk of a low-mass star (K5 or later). The linear extension of S1, if member of the gamma Vel cluster at 370 pc, is $\sim$20\,000 AU. The shape of S1 agrees with a central object and two outflows, and the size is alike the proplyds with negative spectral index detected by \citet{2002ApJ...571..366M}, supporting the association of \#78 (hereafter HHC2008$-$78) with S1.

The source S2 shows three maxima at both frequencies, and an average spectral index $\alpha$ of $-2.1\pm0.4$.  The size of S2, at 370 pc, is 30\,000 AU. The object \#106 from \citet{2008ApJ...686.1195H}, hereafter HHC2008$-$106, superposed to S2, is associated with a proplyd of a star with photometric colors between K5 and M0 and identified as an optically thick disk. 

The source S3 has no counterpart at the consulted databases and a negative spectral index, close to 0.

The source S4 presents a negative spectral index ($-1.8\pm0.7$) and appears as double, with no counterpart in the literature.

The system WR\,11 is located at the centre of S5. The Gamma Vel cluster, formed by low mass stars and protostars \citep{2014A&A...563A..94J}, extends around S5. The stars GES J08092860-4720178 and gamma$^1$ Vel are nearby but detached from S5. The spectral index of this source is $+0.85\pm0.15$, as expected from thermal stellar wind(s). The radio source was detected in the AT20G survey (www.atnf.csiro.au/research/AT20G/) at 5, 8 and 19.9 GHz, with fluxes of $27\pm2$, $50\pm3$ and $86\pm4$ mJy. The corresponding spectral indices are 1.3 (between 5 and 8 GHz), and 0.6 (between 8 and 19.9 GHz). \citet{1985MNRAS.216..613J} observed the system WR\,11 with MOST (band centred at 843 MHz)  and measured a flux of $8.2\pm1.0$ mJy. The spectral index between 843 MHz and 1.4 GHz is $0.6\pm0.3$.

The source S6 is {a} double, and 2 arcmin away from WR\,11. Its spectral index is negative ($-0.9\pm0.25$) and three objects are superposed to it. One is a MOST discrete source, MOST0808$-$471, with a flux of 69$\pm$2 mJy \citep{1985MNRAS.216..613J}.
The other two, catalogued as and X-ray and an IR sources have the same coordinates, and thus could be the same object. \citet{2008ApJ...686.1195H} classify the IR source as a diskless YSO (\#158). The spectral index of S6 between 843 MHz and 1.4 GHz is $-0.95\pm0.15$.

The source S7 correlates with the source IRAS 080484$-$4656, and the double galaxy 2MASX J08100010$-$4705059, 0.17 x 0.17 arcmin$^2$ in size \citep[][2MASS]{2006AJ....131.1163S}. S7 shows a negative spectral index ($-0.9\pm0.3$). 

Source S8 has no counterparts neither in NED nor in the Simbad databases. Its spectral index is {less than $-1.5$}.

   \begin{table*}
      \caption[]{Radio fluxes of the main sources in the field of WR\,11.}
      \label{sources}
     \centering                          
     \begin{tabular}{c r r r r r l l l}        
     \hline\hline                 
ID & RA (J2000) & Dec (J2000) & $S_{\rm 1.4GHz}$ & $S_{\rm 2.5GHz}$ & $\alpha$  & Simbad or NED & Ref. & Notes \\ 
    &  (hms)         &   (dms)         & (mJy)                       &    (mJy)                   &                  &                 counterparts & &   \\
\hline                        
S1 & 08:08:08.409 & $-$47:32:23.08  & 140$\pm$5 & 100$\pm$4 & $-0.6\pm0.14$  &GES J08080881$-$4732336 & (a) & Proplyd \\
  &  &  &  &  &  & GES J08080943$-$4732250 & (a)  & \\
  &  &  &  &  &   &      HHC2008$-$78       & (b) \\
S2 & 08:08:21.330 & $-$47:11:00.63  & 64$\pm$2 & 18$\pm$3   & $-2.1\pm0.4$ & HHC2008$-$106 & (b) & Proplyd \\
S3 & 08:08:23.827 & $-$47:04:21.48  &77$\pm$2  &  70$\pm$4  & $-0.15\pm0.15$ &  &  &  ? \\
S4 & 08:08:45.737 & $-$47:32:29.48  & 12$\pm$1 &    4$\pm$1 & $-1.8\pm0.7$ & &    & ? \\
S5 & 08:09:31.955 & $-$47:20:11.80  & 11.$\pm$0.5 & 18.$\pm$0.5 & $+0.85\pm0.15$ & WR\,11  & & WR\,11\\
S6 & 08:09:42.971 & $-$47:19:39.74  &43$\pm$2 &  25$\pm$2 &  $-0.9\pm0.25$  &MOST0808$-$471 & (c) & ? \\
     &                        &                      &                  &                   &  &2XMM J080942.1$-$471952 &  (d) &\\
      &                      &                      &                   &                    & &2MASS J08094219$-$4719526 & (e)& \\
S7 & 08:10:00.152 & $-$47:05:05.60  & 19$\pm$1 & 11.5$\pm$1 &  $-0.85\pm0.3 $ &IRAS 08084$-$4656 & (f) & R.Galaxy\\
      &        &   &    &    &   &  2MASX J08100010$-$4705059 & (e) &\\
S8 & 08:10:55.213 & $-$47:28:57.68  & 9$\pm$1 & $<3$ & $<-1.5$ & &  & ?\\
\hline  
\end{tabular}
\tabnote{Strongest sources at 1.4 GHz. ID: name given here, coordinates, fluxes measured at L and S bands (see Sect. 3), corresponding spectral index $\alpha$ ($S_\nu \propto \nu^\alpha$), objects from web databases overlapping the sources S1 to S8 together with their references, and most possible counterpart according to this work. (a): \citet{2014A&A...563A..94J}. (b): \citet{2008ApJ...686.1195H}. (c) \citet{1985MNRAS.216..613J}. (d): \citet{2009A&A...493..339W}. (e): \citet{2006AJ....131.1163S}. (f): \citet{1988iras....7.....H}. 
Deconvolved major and minor axis from Gaussian fits are $7.1'' \times 4.7''$ for S3, $4.1'' \times 1.7''$ for S5, and $7.3'' \times 2.5''$ for S7.}
        \end{table*}

   \begin{figure*}[!htb]
\minipage{0.45\textwidth}
  \includegraphics[width=\linewidth]{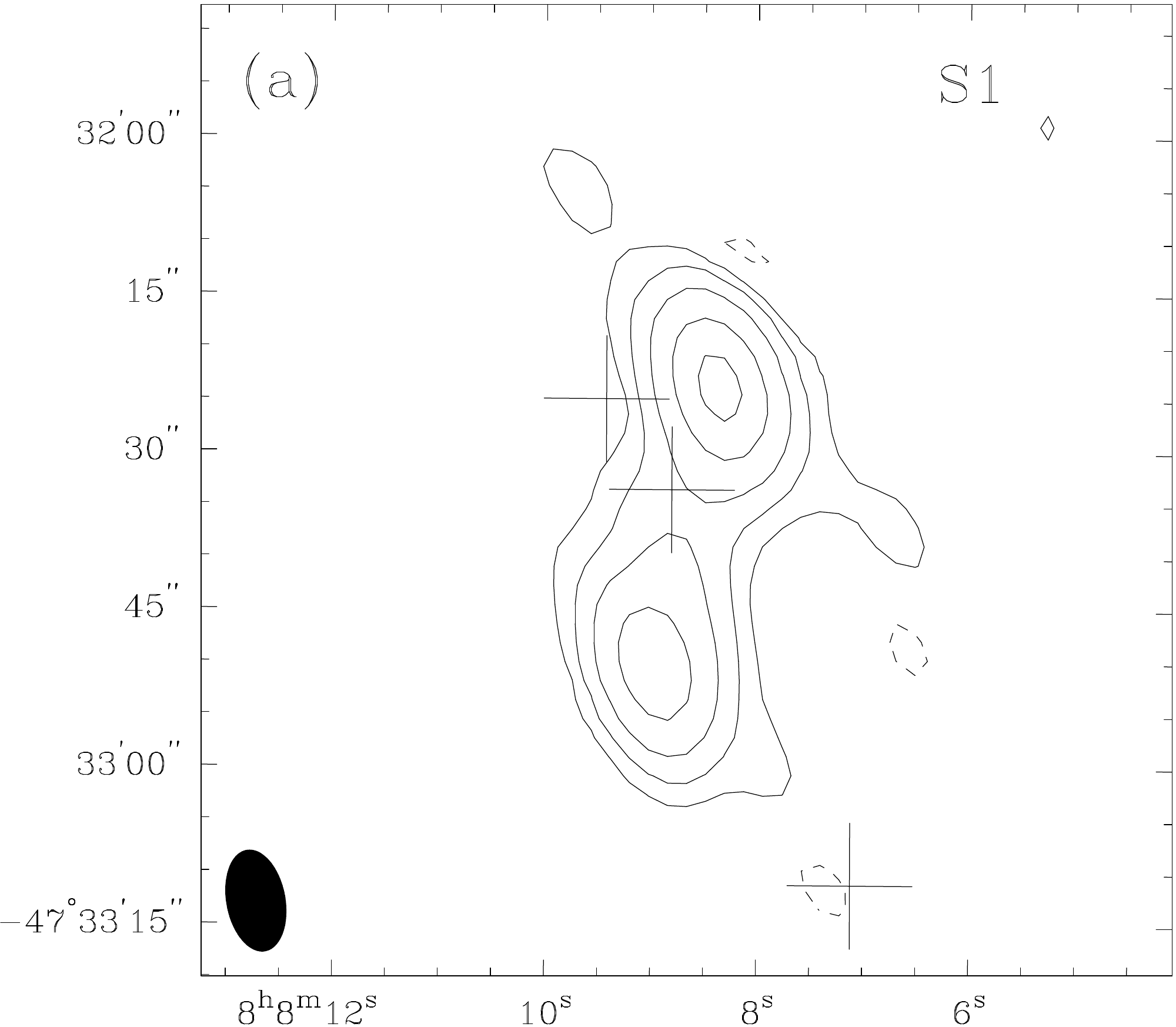}
\endminipage\hfill
\minipage{0.45\textwidth}
  \includegraphics[width=\linewidth]{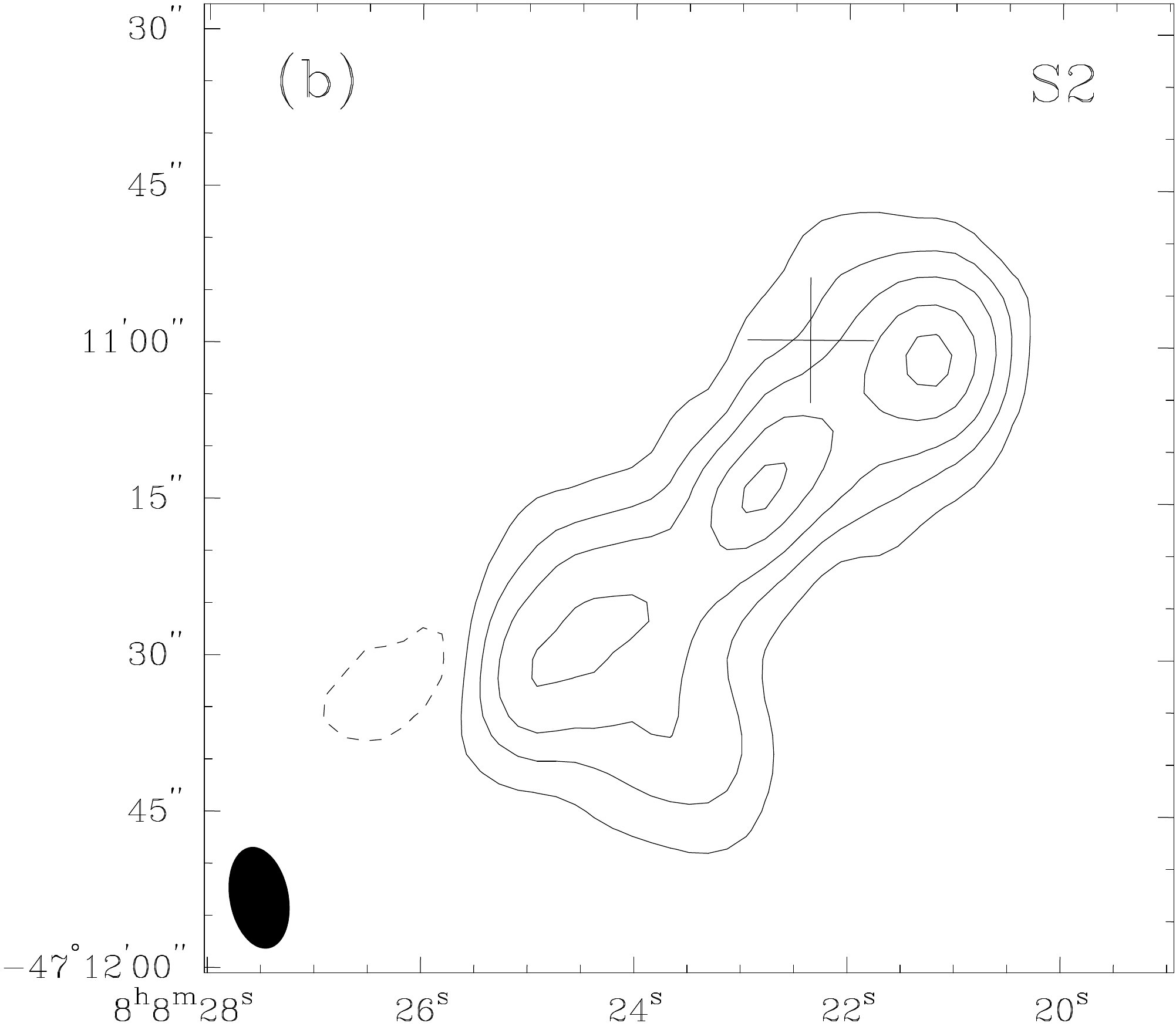}
\endminipage\hfill
\minipage{0.45\textwidth}%
  \includegraphics[width=\linewidth]{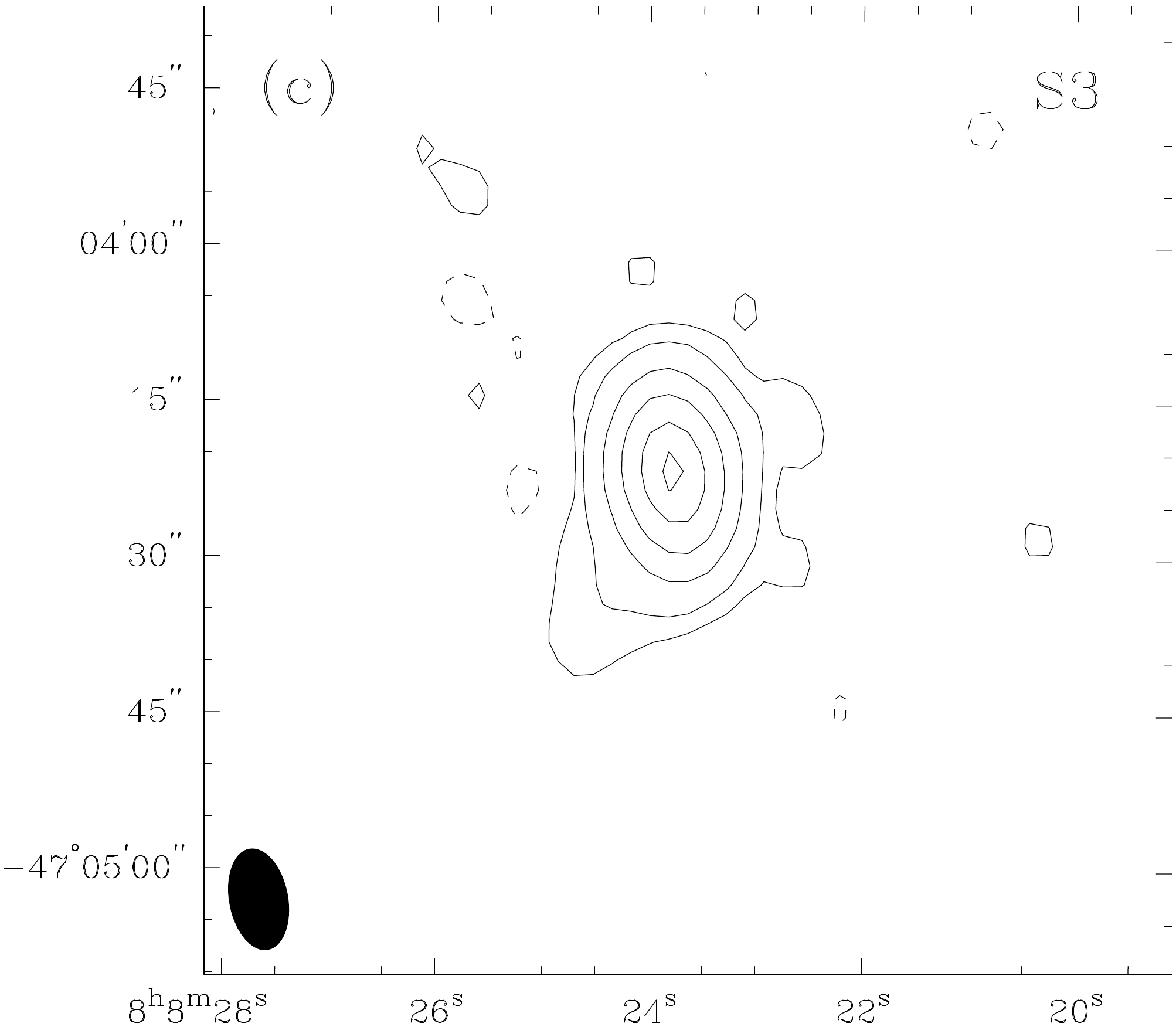}
\endminipage\hfill
\minipage{0.45\textwidth}%
  \includegraphics[width=\linewidth]{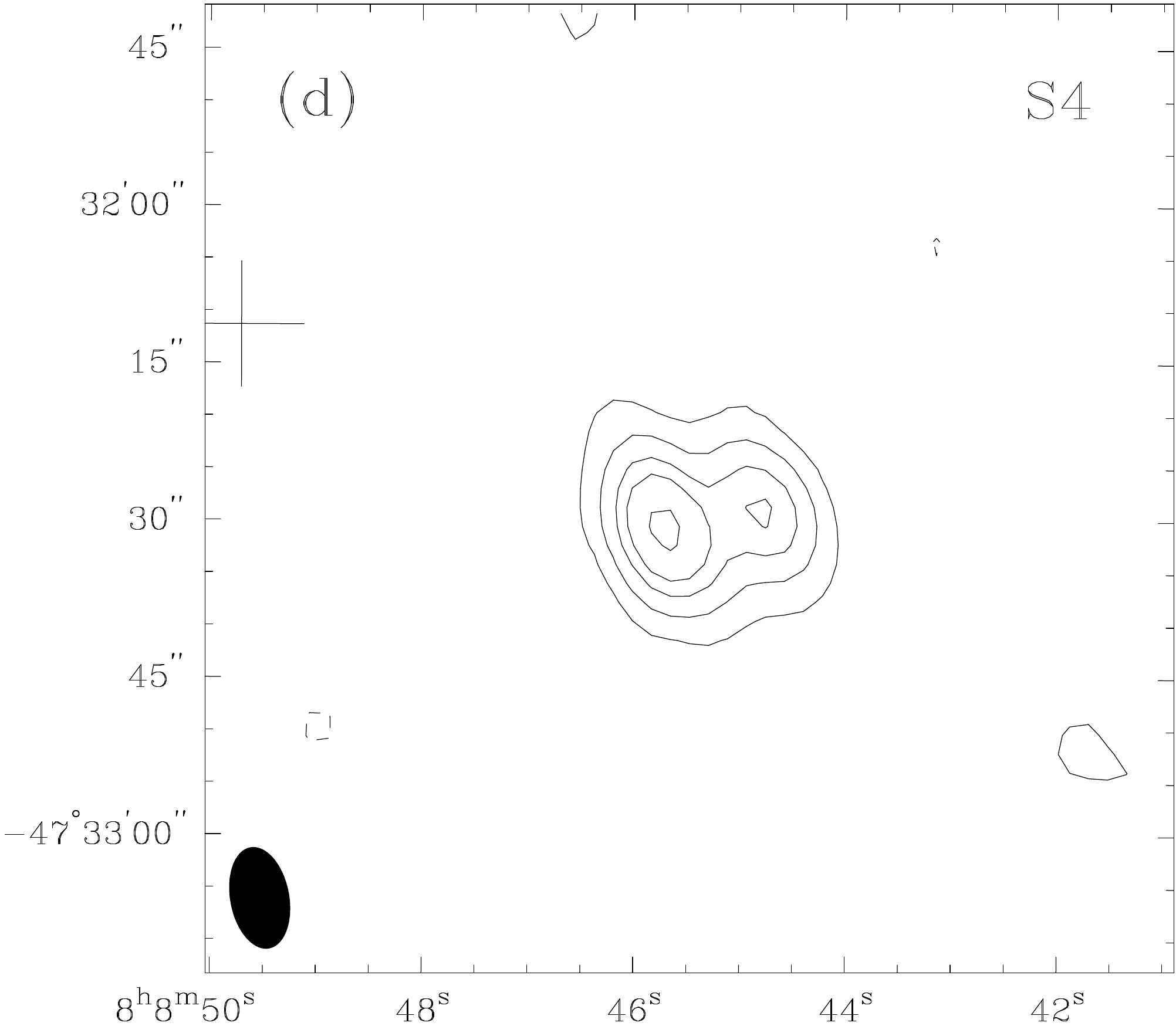}
\endminipage
 \caption{Zoom on sources S1 to S4 (images from C599 data at 1.4 GHz). {Coordinates: RA, Dec (J2000).} Crosses: YSO or protostellar candidates; small boxes: X-ray sources. Black circles: stars.
 (a): S1; contours -3, 3, 10, 30, 90 and 200 in units of $\sigma = 0.2$ mJy beam$^{-1}$.  
 (b): S2; contours -3, 3, 10, 20, 35 and 50 in units of $\sigma = 0.1$ mJy beam$^{-1}$.  
 (c): S3; contours -3, 3, 10, 40, 100, 200 and 300 in units of $\sigma = 0.15$ mJy beam$^{-1}$.  
 (d): S4; contours -3, 3, 10, 20, 30 and 50 in units of $\sigma = 0.09$ mJy beam$^{-1}$.  
 }
\end{figure*}
%
   \begin{figure*}[!htb]
\minipage{0.45\textwidth}
  \includegraphics[width=\linewidth]{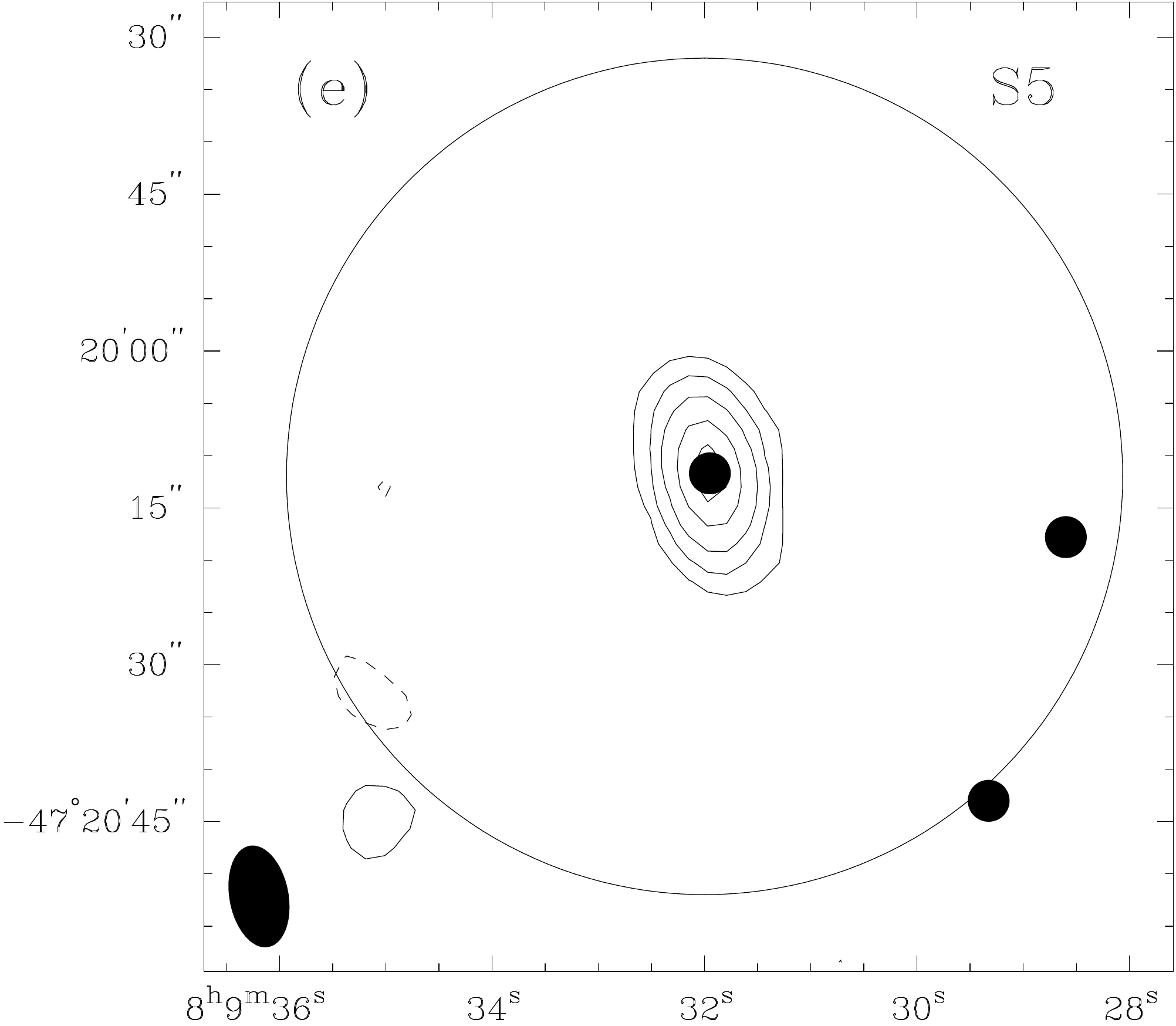}
\endminipage\hfill
\minipage{0.45\textwidth}
  \includegraphics[width=\linewidth]{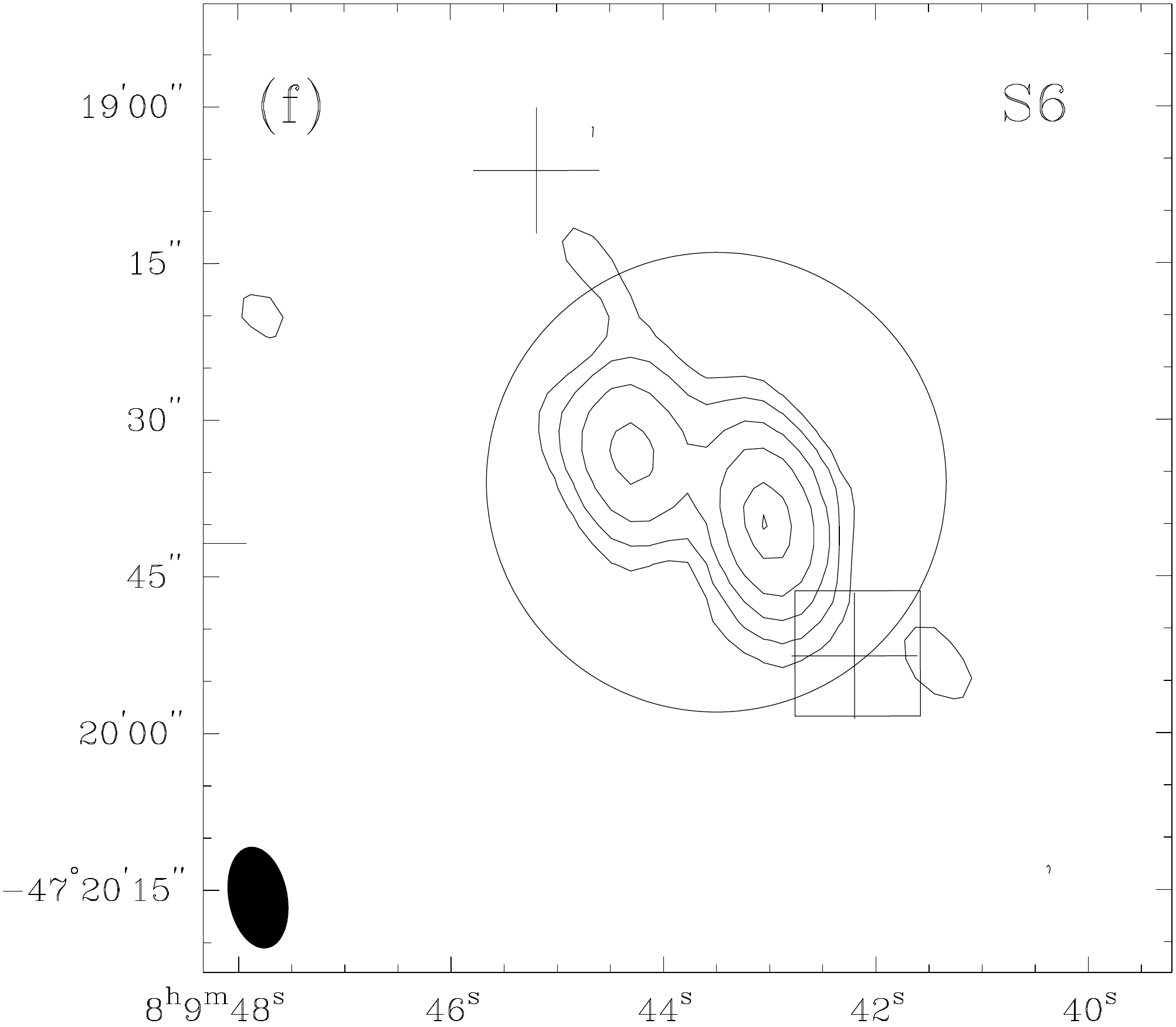}
\endminipage\hfil
\minipage{0.45\textwidth}%
  \includegraphics[width=\linewidth]{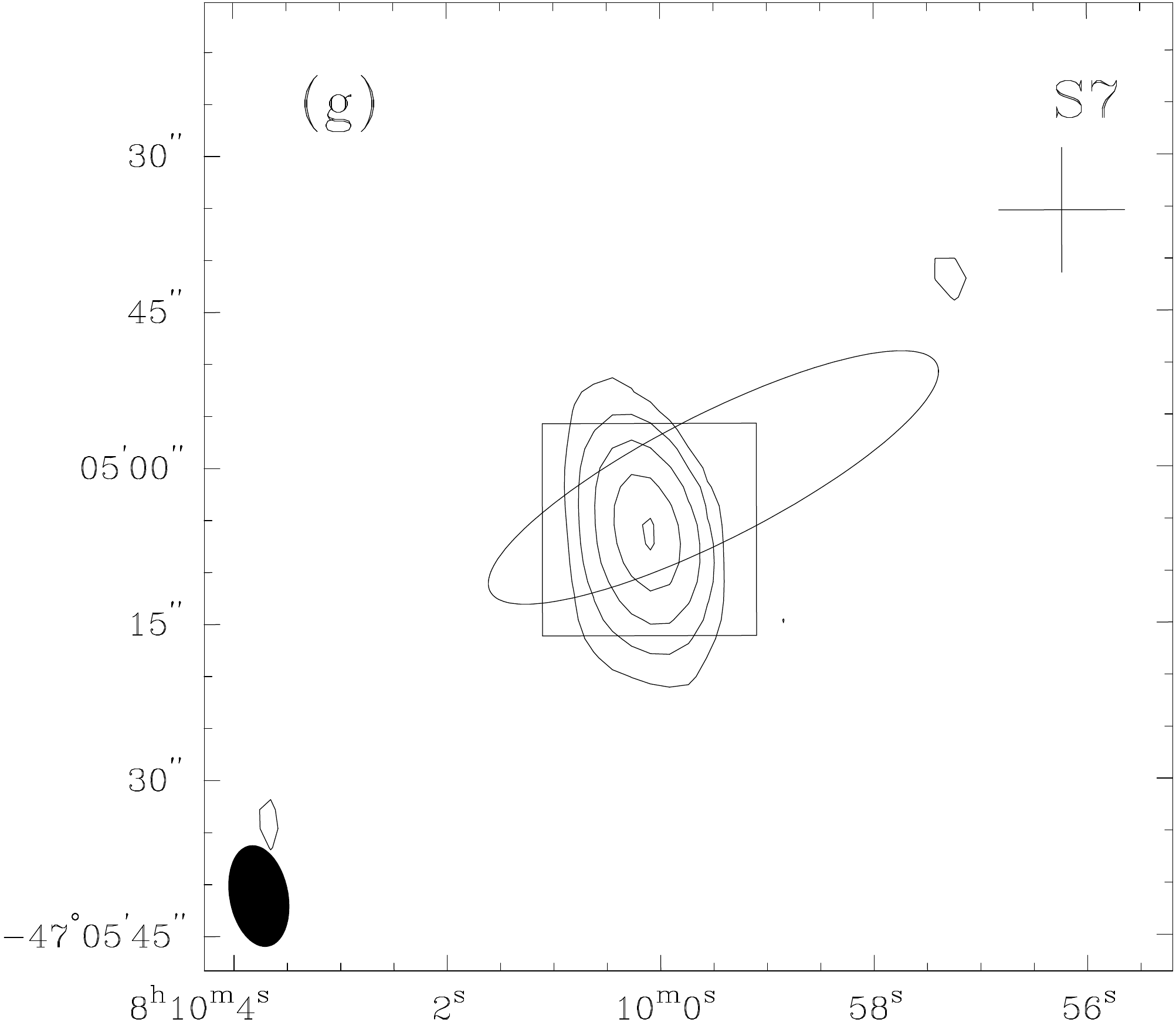}
\endminipage\hfill
\minipage{0.45\textwidth}%
  \includegraphics[width=\linewidth]{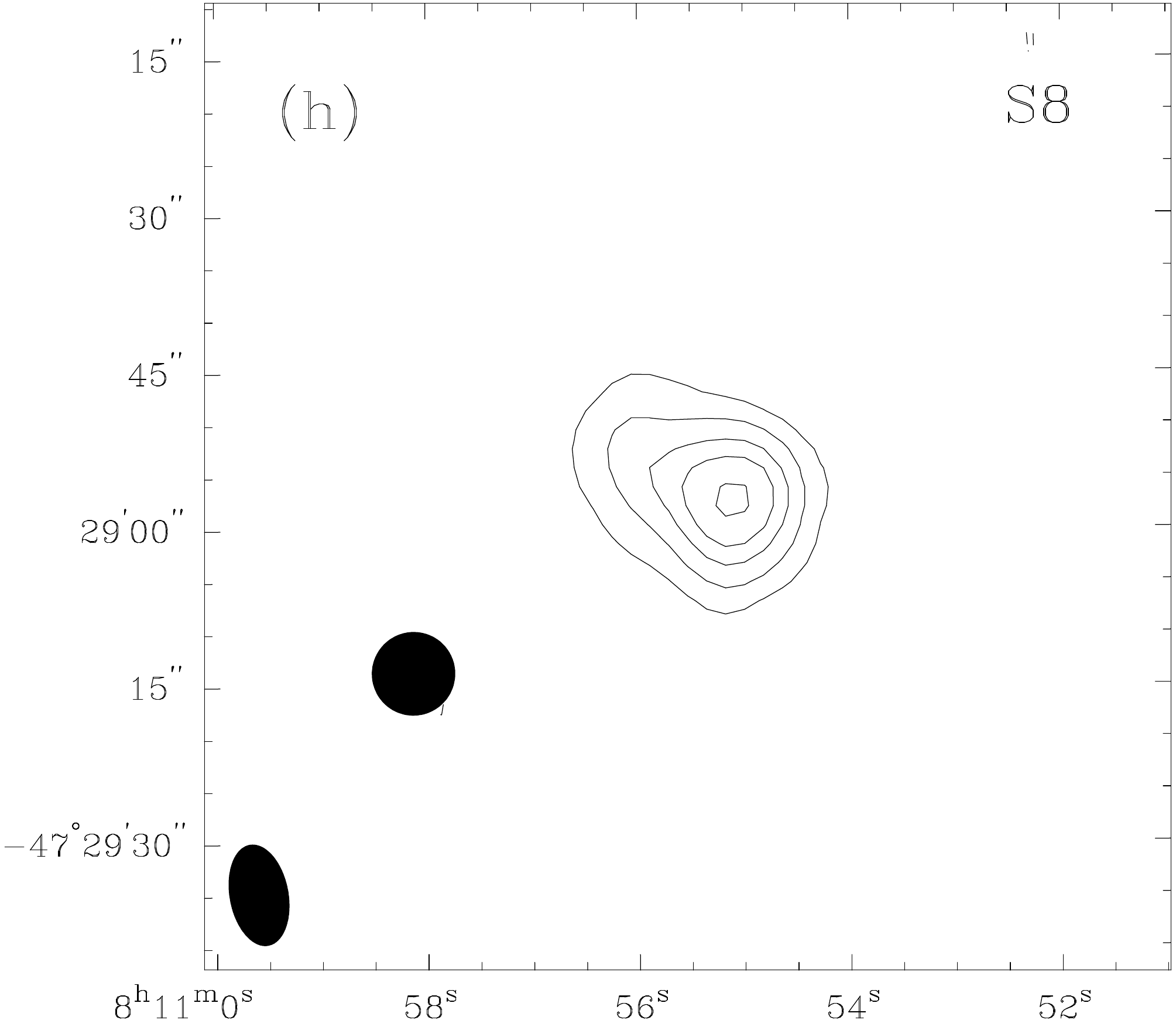}
 \endminipage
 \caption{Same as Fig. 2 on sources S5 to S8 (images from C599 data at 1.4 GHz). 
 (e): S5; contours -3, 3, 10, 25, 55 and 90 in units of $\sigma = 0.09$ mJy beam$^{-1}$. The circle represents the center of the gamma Vel cluster.  
 (f): S6; contours -3, 3, 10, 30, 8, 200 and 280 in units of $\sigma = 0.09$ mJy beam$^{-1}$. 
 (g): S7; contours -3, 3, 10, 40, 80 and 130 in units of $\sigma = 0.09$ mJy beam$^{-1}$. The ellipse marks the IRAS source and the box represents the radio galaxy, see Table 1.   
 (h): S8; contours -3, 3, 10, 20, 30 and 43 in units of $\sigma = 0.09$ mJy beam$^{-1}$.}
\end{figure*}
%

   \begin{table*}
      \caption[]{Sources detected at the 1384-MHz image central region.}
      \label{allsou}
     \centering                          
     \begin{tabular}{l r r r r r l l}        
     \hline\hline                 
$\#$ & RA (J2000) & Dec (J2000) & $S_{\rm 1.4GHz}$ & Deconvolved & Simbad & Notes \\ 
        &  (hms)         &   (dms)         &  (mJy)                       & axes $('','')$&          &            \\
\hline                        
1 &  08:07:53.841 & $-$47:34:22.162 & $2.2\pm0.2$ & & -- & point-like\\
2 & 08:08:08.572 & $-$47:32:33.677 & $145\pm15$ & 36.2,  6.1 & Y*? & discrete\\
3 & 08:08:23.778 & $-$47:04:21.643 & $79\pm2$ & 9.0, 5.0 & --&  discrete\\
4 &  08:08:25.195 &  $-$47:09:53.526 & $1.0\pm0.1$ & 20.1, 10.6 & -- & discrete\\
5 &  08:08:28.339 & $-$47:09:56.829 & $1.5\pm 0.1$ & 23.0, 12.7 & Y*? & discrete\\
6 & 08:08:31.154 & $-$47:10:31.793 & $1.9\pm0.1$ & 12.9, 7.5 & --  & discrete\\
7 & 08:08:42.936 & $-$47:18:37.185 & $1.7\pm0.1$ & & -- & point-like \\
8 &  08:08:45.357 & $-$47:19:28.579 & $1.8\pm0.1$ & 9.6, 2.5 & XMM* & discrete \\
9 &  08:08:45.389 & $-$47:32:29.906 & $12\pm0.5$ & 10.7,  5.8 & -- &  discrete \\
10 & 08:08:55.876 & $-$47:35:03.080 & $5.0\pm0.5$ & & Y*? & point-like \\
11 & 08:09:01.145 & $-$47:18:18.947 & $0.8\pm0.1$ & & --& point-like \\
12 &  08:09:15.986 &  $-$47:09:17.010 & $ 1.2\pm0.1$ &13.2, 8.4 &-- & discrete \\
13 & 08:09:31.956 & $-$47:20:11.610 & $11.3\pm 0.1$ & 6.2, 3.0 &WR\,11 & discrete \\
14 & 08:09:32.971 & $-$47:34:32.881 & $1.5\pm 0.1$  &  14.0,  9.7  & --  &  discrete \\
15 & 08:09:43.112 & $-$47:14:24.504 & $2.0\pm0.1$ & & -- & point-like \\
16 & 08:09:43.318 & $-$47:19:38.027 & $43.0\pm 0.1$ &  16.2, 2.8 & MOST& discrete \\
17 & 08:09:43.748 & $-$47:33:27.319 & $0.7\pm 0.1$ & & Y*? &point-like \\
18 & 08:09:50.561 & $-$47:26:10.844 & $2.7\pm 0.2$ & & pr? &point-like \\
19 &  08:09:51.178 & $-$47:35:46.938 & $7.3 \pm 0.1$ &  7.9,  5.6 & Y*? & discrete\\
20 & 08:09:51.306 & $-$47:27:36.077 & $1.4\pm 0.1$ & & XMM* &point-like \\
21 & 08:09:54.667 & $-$47:30:31.099 & $3.9\pm 0.3$ & 9.4, 1.5 & --& discrete \\
22 & 08:09:55.642 & $-$47:31:29.674 & $1.7 \pm 0.1$ & &-- & point-like \\
23 & 08:10:00.124 & $-$47:05:06.390 & $19.0\pm1.0$ &8.7, 1.5 & IRAS & discrete\\
24 & 08:10:02.680 & $-$47:11:30.112 & $1.7\pm0.1$ & & --& point-like \\
25 & 08:10:13.332 & $-$47:33:55.950 & $1.0\pm0.1$ & & --& point-like \\
26 & 08:10:16.947 & $-$47:19:10.242 & $2.3\pm0.1$ &9.5, 3.2 & --& discrete\\
27 & 08:10:18.286 & $-$47:23:50.957 & $1.3\pm0.1$ &&--& point-like\\
28 & 08:10:23.525 & $-$47:21:11.805 & $1.0\pm0.1$ & &XMM*& point-like \\
29 & 08:10:31.408 & $-$47:28:21.710 & $1.0\pm0.1$ &&--& point-like \\
30 & 08:10:37.417 & $-$47:18:00.427 & $1.0\pm0.1$ &&--&point-like \\
31 & 08:10:45.544 & $-$47:18:52.399 & $1.0\pm0.1$ & &-- & point-like \\
32 & 08:10:50.596 & $-$47:09:26.975 & $1.4\pm0.1$ & & --& point-like \\
33 & 08:10:55.318 & $-$47:28:56.504 & $9.9\pm0.5$ & 12.3, 7.1 &-- & discrete \\
34 & 08:10:56.121 & $-$47:20:12.757 & $0.9\pm0.1$ & 16.1, 6.3 &XMM*& discrete\\
35 & 08:11:03.867 & $-$47:20:07.978 & $3.8\pm0.3$ & 8.1, 3.4 & --& discrete \\
\hline  
\end{tabular}
\tabnote{Y*?: young stellar object candidate; XMM*: 2XMM star; MOST: MOST source; pr?: pre-main sequence star candidate, IRAS: IRAS source. Source \#3 is S3, source \#13 is S5 and source \#23 is S7, see Table 1.
}
        \end{table*}

\section{DISCUSSION}

The search for counterparts of the radio sources detected in this work resulted in possible associations of S1 and S2 with YSO proplyds, S5 with the system WR\,11 and S7 with a double galaxy. 
S6, at $2'$ from WR\,11, is a double source with non-thermal radiation. S4 is alike but weaker and at $\sim 16'$ from WR\,11.
All detected radio sources but S5 show {negative} spectral indices.

The simplest hypothesis is that a single source is responsible of the excess emission detected with Fermi-LAT data. Under such assumption, the two possible scenarios to analyze are whether WR 11, or another source in the field, are the radio counterpart of the Fermi source.

\subsection{Sources other than WR\,11 as the Fermi LAT counterparts}
 
The positional coincidence between the source S1 and the object HHC2008$-$78, and the fact that other proplyds have been detected as radio sources at low frequencies \citep[][and references therein]{2014ApJ...797...60M} favours the identification of S1 with a proplyd. The same can be said of S2 and HHC2008$-$106. Proplyds were detected at more than one frequency, and some of them show a spectral index different from thermal \citep[e.g.][]{2002ApJ...571..366M}. However, none has been identified with (Fermi) gamma-ray sources
\citep{2015ApJS..218...23A,2015arXiv150906382A}.

 The  2MASS catalog of extended sources \citep[XSC in][]{2006AJ....131.1163S} suggests that 2MASX J08100010$-$4705059 is a double-lobed galaxy, based on the emission at J, K and H$_{\rm s}$ bands. The object would also be detected as the source IRAS 08084$-$4656. They are positionally coincident with the radio source S7. There are many Fermi LAT sources identified with radio galaxies, with {gamma-ray} fluxes of the order of the {one} detected by \citet{2015arXiv151003885P}. 
{We note that while the gamma ray fluxes have the same magnitude, the radio flux densities of this source is much lower than those of the Fermi confirmed radio galaxy associations.}
 Unfortunately, since S7 is a discrete source, it is not possible to derive constraints from the radio data presented here.
 
 The source S6 is double, extended along $\sim 45''$. If it is a galactic object (for instance, a microquasar), it should be detected as a stellar source too. The shape and the spectral index are consistent with those of a radio galaxy. Such object, for instance if 30-kpc long, will be at a distance of 100 Mpc, like other typical FRI or FRII galaxies. The Fermi LAT 
energy flux of $1.8\pm0.6 \times 10^9$ erg cm$^{-2}$ s$^{-1}$ quoted by \citet{2015arXiv151003885P} implies, at 100 Mpc,  a  luminosity of $\sim10^{42}$ erg s$^{-1}$. 

A radio emission spectral index of $\alpha = -0.9$ yields an electron energy distribution power law of index $p = 1 - 2\alpha = 2.8$,  and thus a photon spectral index $\Gamma \sim (p+1)/2 = 1.8$ \citep{lapis2008cap1}. 
{The flux measured here corresponds to the extended emission. For most gamma-ray radio galaxies, the major contribution to the gamma-ray emission comes from their inner jets \citep[luminosities and gamma spectral indices of FRI and II found from Fermi LAT data are given in][her Fig. 1]{2012IJMPS...8...25G}. S6 could be a radio galaxy, that radiates} the flux detected by Fermi LAT. And S6 is, as WR\,11, inside the maximum probability contour of the Fermi LAT source position.

 \subsection{WR\,11 as the Fermi LAT counterpart}
 
The detection of a MOST source \citep{1985MNRAS.216..613J} at the position of WR\,11 allows to derive an average spectral index of $+0.6$ from 0.843 to 1.4 GHz, and similar values up to 20 GHz. That NT emission is only present below 843 MHz seems unlikely \citep{2014bApJ...789...87R,2015A&A...577A.100R}. However, absence of NT radio emission is not an argument to reject the hypothesis of a potential association between WR\,11 and the Fermi LAT source. A very low magnetic field or too short energy-loss timescales for electrons would inhibit synchrotron emission to be detected. This emission could be affected by a strong free-free absorption from a dense  circumstellar matter, a small orbit size and/or a complex geometry of the system, too. 

The star Eta Carinae (HD\,93308) shares some features with WR\,11.  The wind kinetic powers are of the same order {\citep[2.8 and 1.6 $\times 10^{37}$ erg s$^{-1}$ respectively,][and references therein]{2013A&A...558A..28D}}. The former is a LVB star with a massive unseen companion \citep{2008MNRAS.384.1649D}, thus considered a CWB. Its radio emission at high-angular resolution was measured since the first times of the ATCA. \citet{2003MNRAS.338..425D} \citep[see also][]{2005ASPC..332..126W} pictured the radio source at 3-cm wavelength as variable between two states: one up to $4'$ in extent, and lower flux ($\sim1$ mJy) at the apastron, and another of a discrete source (extension less than $1''$) and stronger (2 to 3 mJy) near periastron.  The variability correlates with the binary period. The authors interpreted that the orbit of the companion is within an extended and dense medium that is being released by, and largely circumvents, the LBV.
There is no signature that the variability is due to NT radio radiation.

Since Eta Car displays gamma-ray emission \citep{2009ApJ...698L.142T,2010ApJ...723..649A} and NT X-ray emission \citep{2009PASJ...61..629S} it was included in the PACWBs catalogue \citep{2013A&A...558A..28D}.
Very recently, \citet{2015MNRAS.449L.132O} presented a time-dependent model of the HE emission in Eta Car. The model achieved a good agreement with the stellar lightcurve. They found that the measured spectrum can result from accelerated protons that interact with the dense LBV wind: no emission coming from primary electrons is further needed. 
The gamma-ray detectability with current instruments in Ohm et al.$’$ model depends strongly on the stellar mass loss rates. 

WR\,11 is the second CWB with no conclusive signs of NT emission. Its mass loss rate is less than 10 times that of Eta Car \citep[see][and references therein]{2013A&A...558A..28D}; the primary is a WR star instead of a LBV. It was mapped as a discrete source so far (less than $1''$), with no circumstellar emission. A model tailored to this system is crucial to find whether the WR stellar wind is capable of absorbing the radio NT radiation in gamma$^2$ Velorum, and which are the processes that could give rise to the GeV excess of emission.


\section{SUMMARY}

{Telescope data archives are a profitable source of unpublished results.} The reduction and analysis of archival (ATCA) radio interferometric raw data of the field of WR\,11 at 1.4 and 2.5 GHz led to the following findings.

      \begin{enumerate}
    \item There are tens of  radio sources in the field of WR\,11, unclassified and uncatalogued. One {disclosed} in this work  and labeled S6, in principle, is a candidate to be investigated as a gamma-ray producer.
       \item The systems WR\,11 and Eta Car have some similarities but the differences -especially the mass loss rate values and the presence or absence of dense circumstellar matter- preclude the use of same frameworks.
       \item The emission of WR\,11 from frequencies above 843 MHz, up to 20 GHz, presents thermal spectral indices.
       \item The absence of NT emission is not enough to discard WR\,11 as the one producing the Fermi LAT emission excess.     
   \end{enumerate}
   
   Due to sensitivity issues, the 7-yr Fermi data at the position of WR\,11 cannot be used to search variability that could be locked to the system phase, for instance. But the radio sources in the  field can be investigated in more detail, by carrying out observations at very low frequencies and arcsec resolution. The Giant Metrewave Radio Telescope is the best suited to that aim. 
   Radio flux variability studies of the field sources could give a hint on their nature. Together with these observational results, a model to explain the emission of WR\,11 along the entire electromagnetic spectrum is needed.

\begin{acknowledgements}
The author thanks an anonymous referee, whose comments and suggestions resulted in improving the article. PB is grateful to M. Fern\'andez L\'opez, M. De Becker, S. M. Dougherty and S. del Palacio.
This research has made use of the SIMBAD database,
operated at CDS, Strasbourg, France, and of the  NASA/IPAC Extragalactic Database (NED) which is operated by the Jet Propulsion Laboratory, California Institute of Technology, under contract with the National Aeronautics and Space Administration. This work was supported by the FONCyT project PICT 2012-878. 
\end{acknowledgements}

\nocite*{}
\bibliographystyle{apj}
\bibliography{pbbiblist}

\end{document}